\documentclass[sigconf, nonacm]{acmart}

\AtBeginDocument{%
  }

\setcopyright{none}
\renewcommand\footnotetextcopyrightpermission[1]{}
\begin{document}
\title{Leveraging Large Language Models for Generating Labeled Mineral Site Record Linkage Data}

\author{Jiyoon Pyo}
\email{pyo00005@umn.edu}
\orcid{1234-5678-9012}
\affiliation{%
  \institution{University of Minnesota}
  \city{Minneapolis}
  \state{Minnesota}
  \country{USA}
}

\author{Yao-Yi Chiang}
\email{yaoyi@umn.edu}
\affiliation{%
  \institution{University of Minnesota}
  \city{Minneapolis}
  \state{Minnesota}
  \country{USA}
}

\renewcommand{\shortauthors}{Jiyoon Pyo and Yao-Yi Chiang}

\begin{abstract}
Record linkage integrates diverse data sources by identifying records that refer to the same entity. In the context of mineral site records, accurate record linkage is crucial for identifying and mapping mineral deposits. Properly linking records that refer to the same mineral deposit helps define the spatial coverage of mineral areas, benefiting resource identification and site data archiving. Mineral site record linkage falls under the spatial record linkage category since the records contain information about the physical locations and non-spatial attributes in a tabular format. The task is particularly challenging due to the heterogeneity and vast scale of the data. While prior research employs pre-trained discriminative language models (PLMs) on spatial entity linkage, they often require substantial amounts of curated ground-truth data for fine-tuning. Gathering and creating ground truth data is both time-consuming and costly. Therefore, such approaches are not always feasible in real-world scenarios where gold-standard data are unavailable. Although large generative language models (LLMs) have shown promising results in various natural language processing tasks, including record linkage, their high inference time and resource demand present challenges. We propose a method that leverages an LLM to generate training data and fine-tune a PLM to address the training data gap while preserving the efficiency of PLMs. Our approach achieves over 45\% improvement in F1 score for record linkage compared to traditional PLM-based methods using ground truth data while reducing the inference time by nearly 18 times compared to relying on LLMs. Additionally, we offer an automated pipeline that eliminates the need for human intervention, highlighting this approach's potential to overcome record linkage challenges.

\end{abstract}

\ccsdesc[500]{Information systems~Entity resolution}
\ccsdesc[300]{Information systems~Geographic information systems}

\keywords{Spatial entity linkage, entity matching, geospatial data}

\maketitle
\section{Introduction}
The surge of web data has enabled the creation of rich and diverse datasets by merging information collected from various sources, including web pages and research conducted by different individuals. However, this abundance of data has also increased data heterogeneity, emphasizing the need for efficient and effective data merging from various sources. Record linkage, the process of identifying records that refer to the same entity across datasets, is crucial to overcoming inconsistencies and discrepancies between datasets and building a comprehensive database. 

\begin{figure*}
    \centering
    \includegraphics[width=0.9\linewidth]{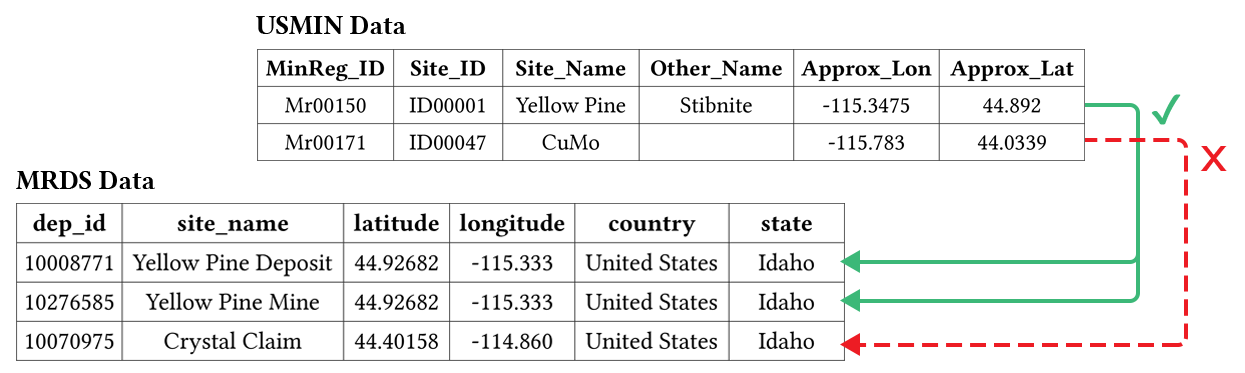}
    \caption{Illustration of the mineral site record linkage process. The pipeline must accurately link records despite variations in recorded information and style. Green highlights records that should be linked, while red indicates records that should not be linked.}
    \label{fig:mineral-site-linkage}
\end{figure*}

Information about mineral sites, such as their current status, available minerals, ownership, site name, and location, is recorded and stored in mineral site databases. In the context of mineral site data (e.g., Mineral Resources Data System (MRDS)~\cite{mrds} and USMIN Mineral Deposit Database~\cite{usmin}), record linkage is critical to identify and map mineral deposit areas accurately. Figure \ref{fig:mineral-site-linkage} demonstrates an example of linking mineral site records between USMIN and MRDS. These databases contain records with missing values (i.e., nulls), attributes that indicate unique identifies (e.g., `Site\_ID' for USMIN and `dep\_id' for MRDS), inconsistent attribute naming conventions (e.g., `Approx\_Lon' for USMIN and `longitude' for MRDS), and variation within attribute values (e.g., `Yellow Pine', `Yellow Pine Deposit', and `Yellow Pine Mine').

\begin{figure*}[htbp]
    \centering
    \includegraphics[width=0.9\linewidth]{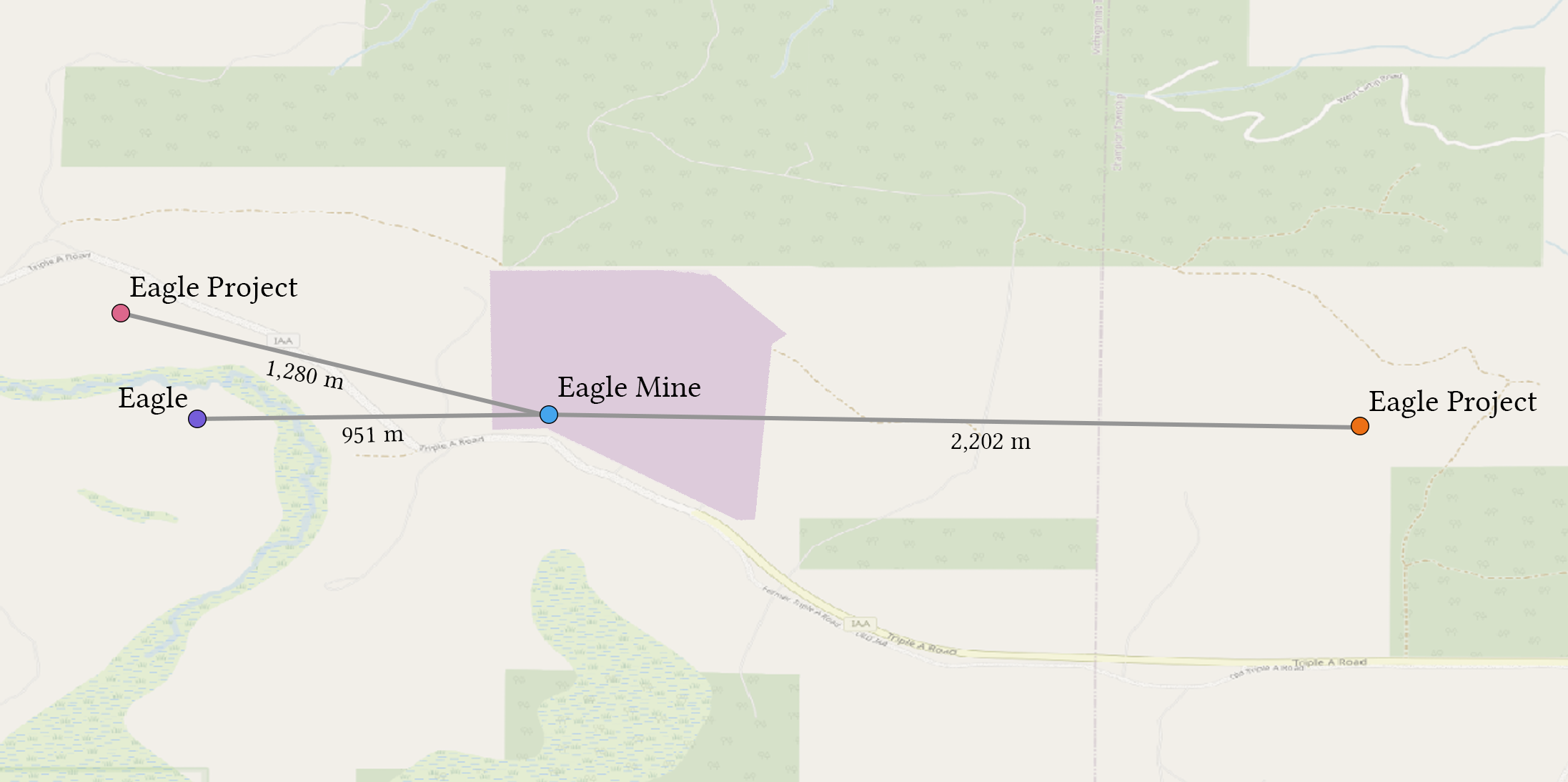}
    \caption{Image of Eagle Mine, where each color represents a mineral site record from different databases~\cite{mrds,usmin,eagle1,eagle2}. The actual polygon region of Eagle Mine is highlighted in light purple~\cite{osm}.}
    \label{fig:eagle-mines}
\end{figure*}

\begin{figure*}[htbp]
    \centering
    \includegraphics[width=0.9\linewidth]{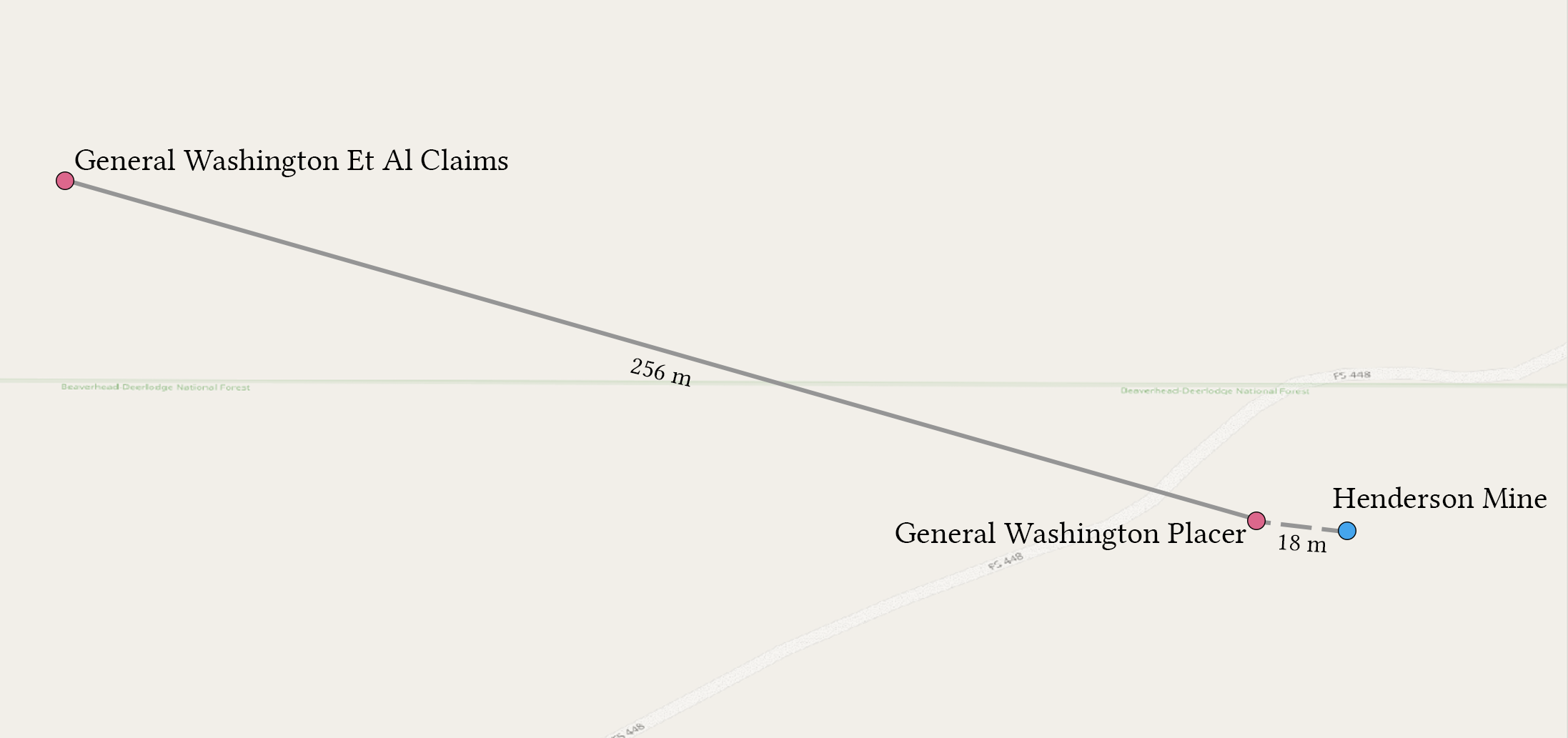}
    \caption{Image of General Washington Placer and Henderson Mine, with mineral records in different colors to represent distinct mineral sites.}
    \label{fig:far-mines}
\end{figure*}

Regarding spatial characteristics, recorders often simplify polygon-shaped location geometries to point data using generalization techniques such as centroid approximation to enhance data comprehension. The level of abstraction varies based on the intended use of the data~\cite{generalization}, but it can significantly distort the spatial relationships between entries. For example, Figure \ref{fig:eagle-mines} illustrates four points representing Eagle Mine in Michigan, along with the actual polygon location of the mine (shown in pink). Although the points are spatially dispersed, they all refer to Eagle Mine and should be linked. In contrast, Figure \ref{fig:far-mines} shows geographically close points representing different mineral sites. An effective linkage model must differentiate these points by recognizing the difference in site names, even if the points are located nearby.

Linking mineral sites poses additional challenges due to the spatial attributes present in each record. First, location information is often ambiguous; for instance, one record may describe the mine entrance, while another may reflect the center of the mineral deposit. Along with that, the geographical location is often unreliable, as some records are inaccurately positioned, ranging up to 50 kilometers\footnote{This issue was noted by Michael Zientek of the United States Geological Survey (USGS) during record verification.} due to older, less precise methods of data collection or a significant map generalization. Furthermore, site names often include the names of the minerals present at the site (e.g., `Crescent Creek \underline{Cr-Pt}' meaning Crescent Creek \underline{Chromium-Platinum}) or deposit types (e.g., `Silver Creek \underline{Placer}'), complicating cross-dataset comparisons. The difficulties mentioned above highlight the need for robust record linkage models to accommodate such discrepancies and ensure a more accurate and comprehensive mapping of mineral resources.

Prior approaches~\cite{ditto, 10.1145/3485447.3512026} to record linkage often rely on pre-trained discriminative language models (PLMs), such as Bidirectional Encoder Representations from Transformers (BERT)~\cite{bert}. PLM-based record linkage approaches typically cast the record linkage problem as a binary classification problem, which requires a large amount of labeled data (i.e., data indicating whether each pair of records represents the same entity) for accurate performance. However, curating ground-truth data is costly, time-consuming, and requires domain knowledge; this makes using PLMs impractical for large-scale record linkage applications. In the scope of mineral site linkage, experts can take up to several days to link datasets consisting of a couple of hundred records\footnote{We have learned the following information from United States Geological Survey experts who have consulted our project. See also~\citet{tungsten-assessment}.}. Additionally, fine-tuned models show a lower performance when there is a significant data imbalance in the training data, where non-matching records are vastly outnumbered by matching ones. When trained on such imbalanced data, PLMs often over-predict \textit{non-match} pairs to boost the overall model accuracy.

Recent advancements in large generative language models (LLMs) have shown promising potentials of improving record linkage~\cite{foundation-wrangle, peeters}, even when they are not trained on domain-specific data (i.e., zero-shot setting). In other words, LLMs can classify whether two records are linked without extensive training data. Pre-trained on an extensive amount of datasets, even beyond the amount used to pre-train discriminative models, LLMs can effectively understand and compare attributes, even when named differently (e.g., `site name' and `other names'). For instance, if an LLM is presented with two records--one listing `Tungsten Jim' as the main mineral site name and the other referring to `Tungsten Jim' as an alternate name--it can accurately link the records without manual rules stating that the two represents the same attribute. This capability arises from LLM's ability to recognize that both records refer to the same mineral site due to their locational proximity and identical naming despite the variation in the attribute label. However, the practical application of LLMs for record linkage is often limited due to their high computational cost and long processing time, posing issues in scalability. For record linkage, LLMs are typically provided with serialized text inputs representing two records~\cite{foundation-wrangle}, where each can contain over 50 attributes. The cost of using LLMs depends on the input length, making their use expensive for the record linkage task. As LLMs generate text token by token, repeated calls are required to generate these sequences, increasing the computational demands. 

In this work, we propose an approach that combines the strengths of both PLM and LLM to optimize the record linkage task for mineral site data. Our method leverages the generative capabilities of LLMs to create high-quality training data, which are used to fine-tune a PLM for a binary classification task (i.e., classifying a pair of records as a \textit{match} or a \textit{non-match}). By utilizing LLMs, our two-stage approach mitigates the challenge associated with obtaining and labeling extensive training data. Additionally, our method maintains the inference time of PLMs, which is nearly 18 times lower than the inference time of LLMs. The proposed approach offers a scalable and effective strategy for addressing the complexities of real-world record linkage problems.

We summarize our contributions as follows:
\begin{itemize}
    \item \textbf{Development of a Hybrid Record Linkage Method}: We introduce an approach that combines the generative capabilities of large language models with the classification efficiency of pretrained language models. 
    \item \textbf{Improvement in Scalability and Efficiency}: By mitigating the challenge of obtaining and labeling extensive ground truth data, our method offers a scalable and efficient solution for large-scale data linkage, making it particularly applicable to complex real-world scenarios, such as record linkage for mineral site data.
    \item \textbf{Analysis of Data Requirements and Data Imbalance}: We conduct comprehensive research to analyze the size of training data required for each class and assess how data imbalance impacts the performance of our model.
\end{itemize}

\section{Preliminaries and Related Work}
In this section, we define record linkage (Section \ref{sec:task-definition}), distinguish PLMs and LLMs (Section \ref{sec:plmllm-definition}), and review related work on utilizing PLMs and LLMs for record linkage (Section \ref{sec:related-works}).

\subsection{Task Definition} \label{sec:task-definition}
We define the record linkage problem as follows:

Given a set of databases, $\{D_{1}, D_{2} \cdots D_{i}\}$, where $D_{j}$ represents a database, the task is to identify and link all records referring to the same entity within and across all available databases. Each entity is characterized by a set of general attributes (e.g., place name, available resources) and spatial attributes, which include point data that could be derived from line or polygon geometries. The linkage task involves matching entities across databases that may have different schemas and missing attributes (i.e., null values). 

\subsection{Definition of PLMs and LLMs} \label{sec:plmllm-definition}
In this work, we define pre-trained discriminative language models, which are encoder-based, as PLMs. Examples of PLMs include BERT~\cite{bert}, RoBERTa~\cite{roberta}, and DistilBERT~\cite{distilbert}. We refer to decoder-based generative language models, including all models developed after GPT-3~\cite{gpt3}, as LLMs. Examples of LLMs include GPT-4~\cite{gpt4o}, LLaMA~\cite{llama}, and Gemma~\cite{gemma}, which are designed for text generation tasks.

\subsection{Related Work}\label{sec:related-works}
Record linkage is the task of integrating data to identify and link records in one or more databases that refer to the same entity. By linking different datasets, record linkage enhances the amount of information available for each record, resulting in a more comprehensive database.

\textbf{Traditional Record Linkage} Earlier approaches~\cite{10.1145/3347146.3359345} often use traditional string similarity metrics, such as Levenshtein distance or Jaccard similarity, to determine whether two records refer to the same entity. These methods rely on empirically defined rules and weights applied to manually selected attributes (e.g., name, social security, or phone number) to calculate the similarity score. To resolve ambiguities, researchers use crowd-sourcing to eliminate confounding factors or add additional details to assist the linking process in their models. As stated, these methods require extensive manual rule development and data processing. Our work aims to develop a fully automated process that minimizes or eliminates the need for human intervention.

\textbf{PLM-based Record Linkage} To enhance the accuracy and scalability of entity linkage, recent approaches shift towards implementing deep learning models, such as Long Short-Term Memory (LSTM)~\cite{lstm} and BERT~\cite{bert}. These approaches not only reduce manual processing but also enhance the accuracy of entity linkage. DeepER~\cite{deeper} introduces a method that uses LSTM networks and Global Vectors (GloVe)~\cite{pennington-etal-2014-glove} text embeddings to capture semantic similarities between records. Similarly, Ditto~\cite{ditto} converts the details of each record to a sentence and uses Sentence-BERT~\cite{sentence-bert} to generate contextualized embeddings for sentences. PLM-based record linkage methods have demonstrated enhanced performance by considering all attributes available in the data rather than focusing solely on manually selected key fields. However, their effectiveness is limited by the availability of training data. \citet{sample-size} demonstrates that PLM-based methods fail to predict a class when there are fewer than 25 instances of that class and require at least 100 labeled examples to achieve a reasonable level of accuracy. Our approach addresses this challenge by generating synthetic training data, which creates a framework that reduces dependency on large volumes of labeled data.

\textbf{LLM-based Record Linkage} There is a growing interest in developing unsupervised or semi-supervised record linkage approaches to minimize the reliance on large labeled datasets. Large language models (LLMs), pre-trained on vast amounts of data, have shown the potential of being a zero-shot solution for record linkage across diverse domains. ~\citet{foundation-wrangle} suggests applying LLMs to this task, achieving state-of-the-art results without fine-tuning the model for record linkage. The researchers use serialization methods to convert structured tabular data into textual inputs to formulate the record linkage problem as a text generation task. ~\citet{peeters} further demonstrate that domain-specific prompting (e.g., asking to identify for the same `mine' instead of `entity') and strict binary responses prompting (i.e., ``Yes'' or ``No'') lead to more stable results and significantly improve LLM's performance. LLMs such as Generative Pretrained Transformer (GPT)~\cite{gpt} and Large Language Model Meta AI (LLaMA)~\cite{llama} hold extensive world knowledge by being trained on large-scale data, offering promises for record linkage, especially in scenarios where labeled training data is scarce or costly. However, their long inference time limits their applicability in record linkage tasks, which often involve processing large databases.

\textit{Blocking} To reduce the number of comparisons in record linkage, a common approach is to use `blocking' to create candidate pairs by identifying records that are most likely to represent a \textit{match}. However, many state-of-the-art blocking methods are unsuitable for our scenario. The earliest deep learning-based method, DeepER~\cite{deeper}, generates candidate pairs by aggregating all values in a record to tuple vectors. This approach cannot be applied to our dataset since the details present in each record differ, with the significance of each attribute value varying. Some columns may be irrelevant, but without proper weighting, all attributes are treated equally, which can distort the results. A more advanced method, AutoBlock~\cite{autoblock}, improves DeepER by learning aggregation weights from labeled data. Yet, the mineral site record linkage case lacks sufficient labeled data to train this method effectively. 

DeepBlock~\cite{deepblock} uses hash-blocking with word embeddings to compute semantic similarity between attribute names. In mineral site linkage, the number of attributes often depends on the granularity of the records. For example, in one data (e.g., USMIN~\cite{usmin}), minerals available at a site might be listed under a single `commodity' column, while in another data set (e.g., MRDS~\cite{mrds}), the same information might be spread across multiple columns (e.g., `commod1', `commod2', `commod3') based on the mineral availability. This inconsistency introduces challenges when applying this blocking method to our approach, highlighting the limitations of blocking methods in the mineral site record linkage context.

\textbf{Spatial Record Linkage} There has been research conducted specifically for spatial record linkage where researchers try to preserve the spatial semantics during the linkage process rather than considering them as plain strings~\cite{magellan}. Existing spatial entity linkage methods~\cite{10.1145/3347146.3359345} often depend on cleaned data that adheres to a predefined schema with latitude and longitude points. These methods typically calculate spatial distances, such as Haversine distance, to determine whether two spatial entities are likely to represent the same entity. For example, GeoER~\cite{10.1145/3485447.3512026} leverages a BERT-based model to compare non-spatial attributes while relying on Haversine distance calculation to compare spatial attributes. SkyEx~\cite{pareto} utilizes spatial blocking to filter out non-match based on the distance between points of interest (POIs). They then use a preference function to rank pairs based on the likelihood of referring to the same entity. 

\begin{figure*}[htp!]
    \centering
    \includegraphics[width=0.9\linewidth]{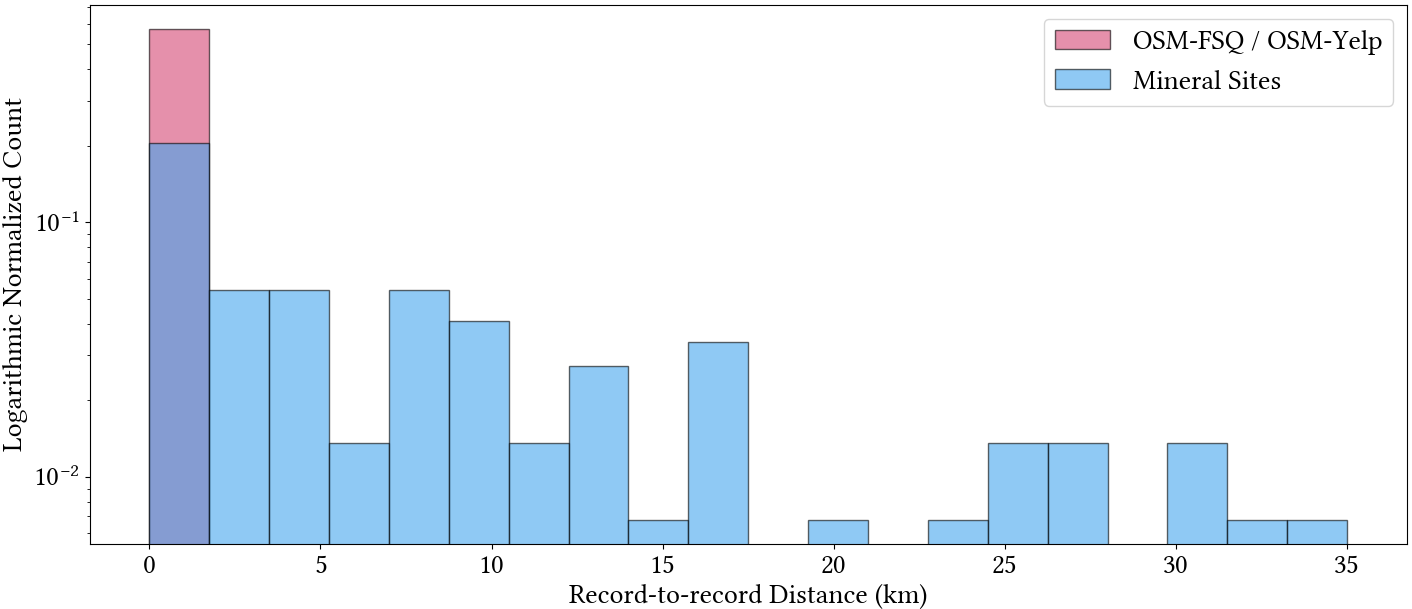}
    \caption{Record-to-record distance distribution of match data from OSM-FSQ/OSM-Yelp compared to mineral sites data. While OSM-FSQ and OSM-Yelp data fits within a 2.5-kilometer range, the distance between match mineral site records can range up to 35 kilometers, with a similar distribution across the bins. This highlights the limitation of spatial record linkage methods that rely on empirically defined distance thresholds since such approaches may not be suitable for domains with large spatial distance variance, like mineral sites.}
    \label{fig:mine-length-comparison}
\end{figure*}

Prior spatial record linkage methods have primarily focused on POI data. Unlike POIs, the spatial distance between two mineral site records can vary significantly depending on the type and location of the deposit, making these previous approaches less effective when linking mineral site records. As shown in Figure \ref{fig:mine-length-comparison}, while match records in the OSM-Foursquare (FSQ) and OSM-Yelp datasets have distance proximity within a 2.5-kilometer range, match records in mineral site databases can have distances spanning up to 35 kilometers, with comparable amounts of data in each distance range. Due to this variability, we cannot rely on predefined distance thresholds and models trained on POI data for our problem. For example, GeoER~\cite{10.1145/3485447.3512026} uses a 2-kilometer boundary in the initial step to reduce the number of comparisons. However, when linking mineral site records, selecting an appropriate distance threshold becomes ambiguous due to the significant variability in distances, indicating that such methods cannot be adapted to our context. 

While traditional and advanced methods offer valuable insights, they often require extensive manual effort and substantial training data or are computationally expensive to be scaled to a large dataset. Our proposed approach addresses these limitations by utilizing LLMs to generate training data and fine-tune PLMs, thereby enhancing record linkage performance in scenarios with limited labeled data.

\section{Proposed Approach}
Our proposed record linkage method leverages the extensive knowledge of LLMs and the efficient inference capabilities of PLMs. The methodology comprises two main steps: (1) generating labeled training data using LLMs and (2) performing record linkage using a PLM. This two-step approach addresses the challenge of limited ground-truth training data and reduces the long inference times associated with relying solely on LLMs for the entire record linkage process.

\subsection{Training Data Generation}
To generate the training data, we use LLaMA3-8b~\cite{dubey2024llama3herdmodels}, an open-source LLM. While larger and more recent models, such as GPT4o~\cite{gpt4o}, may potentially offer better performance, we selected LLaMA3-8b due to computational constraints and the potential for fine-tuning the model to create more precise labels. Using this model allows for future optimization without overburdening available resources. Using LLMs as a method to generate training data involves framing the record linkage problem as a text-generation task. Therefore, we use the tabular data serialization method proposed by ~\citet{foundation-wrangle}. Each entity's attributes and values are serialized as follows:

\begin{center}
    \verb|serialize(e):=attr|\textsubscript{$1$}\verb|:val|\textsubscript{$1$}\verb|...attr|\textsubscript{$i$}\verb|:val|\textsubscript{$i$}
\end{center}

For record pair serialization, we follow the method proposed by \citet{peeters}, prompting the LLM to label the record pairs using the following template:

\begin{center}
\vspace{0.1em}
\noindent\fbox{%
    \parbox{0.97\linewidth}{%
    {\fontfamily{cmtt}\selectfont 
        Entity A is serialize(A). \\
        Entity B is serialize(B). \\
        Do the two mine descriptions refer to the same real-world mine. Answer with `Yes' if they do and `No' if they do not. \\
        Answer only in Yes or No.}
    }%
}%
\end{center}

Previous study~\cite{peeters} demonstrates that restricting responses to ``Yes'' or ``No'' improves the performance of LLMs. Without explicit prompts enforcing strict response formats (i.e., `\verb|Answer| \verb|only| \verb|in| \verb|Yes| \verb|or| \verb|No.|'), LLMs tend to generate extended reasoning. To eliminate the need for manual post-processing, we limit the LLM's response to ``Yes'' or ``No''.

We convert the ``Yes'' or ``No'' labels generated by the LLMs into binary values, with 1 representing a \textit{match} and 0 representing a \textit{non-match}. This creates a dataset that follows the structure shown in Table \ref{tab:sturcutre}. The columns uri\_1 and uri\_2 represent the unique ID for each record. In cases where the unique ID is not available, we use the row number as the unique identifier.

\begin{table}[htp!]
    \centering
    \begin{tabular}{|c|c|c|}
    \hline
        uri\_1 & uri\_2 & llama\_prediction \\
        \hline \hline
        $\text{ID}_{\text{Entity A}}$ & $\text{ID}_{\text{Entity B}}$ & 0 \\
        $\text{ID}_{\text{Entity A}}$ & $\text{ID}_{\text{Entity C}}$ & 0 \\
        $\text{ID}_{\text{Entity A}}$ & $\text{ID}_{\text{Entity D}}$ & 1 \\
         \hline
    \end{tabular}
    \caption{Structure of LLM synthesized data.}
    \label{tab:sturcutre}
\end{table}

\subsection{Record Linkage}
For the record linkage task, we utilize RoBERTa~\cite{roberta}, a robustly optimized variant of BERT~\cite{bert}, as the PLM for fine-tuning. We select RoBERTa over BERT since RoBERTa shows better performance across various downstream natural language processing (NLP) tasks than BERT. This improvement is due to RoBERTa's training on a larger dataset, which increased from 16GB to 160GB~\cite{roberta} compared to BERT, resulting in a more extensive vocabulary. Our empirical results, detailed in Appendix \ref{sec:different-bert}, demonstrate that fine-tuning RoBERTa outperforms BERT and other variants, such as DeBERTa~\cite{deberta}.

To evaluate our approach, we fine-tune the model using LLaMA-labeled data. RoBERTa, similar to BERT, provides deeply contextualized embeddings that capture the semantic meaning of the input data, which is crucial for effective record linkage. To prepare the data, we adopt the text serialization approach proposed in Ditto~\cite{ditto} since the suggested method is widely used in PLM-based record linkage tasks (e.g., ~\citet{10.1145/3485447.3512026}). For each data entry, \verb|e=|$($\verb|attr|\textsubscript{$i$},\verb|val|\textsubscript{$i$}$)$\textsubscript{$1\leq i\leq k$}, we serialize the attributes and values as follows:
\begin{center}
    \verb|serialize(e):= [COL]attr|\textsubscript{$1$}\verb| [VAL]val|\textsubscript{$1$}\verb|       |
    
    \verb|                 ... [COL]attr|\textsubscript{$i$}\verb| [VAL]val|\textsubscript{$i$}
\end{center}

\noindent where \verb|[COL]| and \verb|[VAL]| are special tokens indicating the start of attribute names and values, respectively. 

For serializing record pairs \verb|(e|\textsubscript{$1$}\verb|,e|\textsubscript{$2$}\verb|)|, we use the following format: 
\begin{center}
    \verb|serialize(e1,e2):= [CLS] serialize(e|\textsubscript{$1$}\verb|)        | 
    
    \verb|                      [SEP] serialize(e|\textsubscript{$2$}\verb|) [SEP]|
\end{center}
\noindent where \verb|[CLS]| is the special token indicating the beginning of the sequence, and \verb|[SEP]| is the special token separating the two records\footnote{We modify RoBERTa's original special token dictionary, replacing the beginning-of-sequence token <s> with [CLS] and the end-of-sequence token </s> with [SEP].}.

We fine-tune the PLM so that, given a serialized record pair as input, it outputs one of the binary labels, which corresponds to \textit{match} or \textit{non-match}.

\section{Evaluation}
\begin{figure*}
    \centering
    \includegraphics[width=\linewidth]{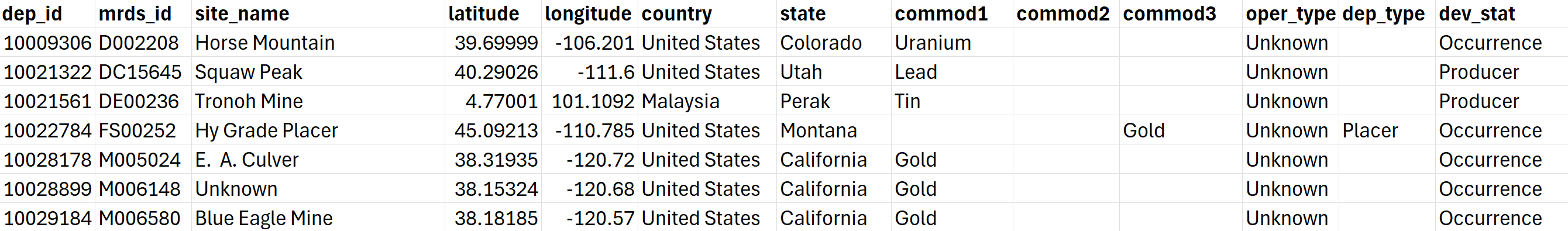}
    \caption{Sample of MRDS Data displaying the heterogeneity of the data. Some of the attributes are left blank, and some attributes consist purely of unique values.}
    \label{fig:mrds}
\end{figure*}

We evaluate the effectiveness of our proposed approach by applying it to mineral sites throughout the United States. For this evaluation, we use the Mineral Resources Data System (MRDS)~\cite{mrds} and the United States Mineral Deposit Database project (USMIN)~\cite{usmin} databases, which are comprehensive mineral site repositories recording various characteristics of mining sites, such as site names, alternative names, geographical locations, and the types of minerals present. The heterogeneity of these databases, in terms of schema differences and data representation, makes them well-suited for testing the robustness and adaptability of our approach. Figure \ref{fig:mrds} displays an example of heterogeneity that is common in the dataset: many of the items are null values, such as `commod2' and `commod3', and some columns only consist of unique items for each record such as `dep\_id' and `mrds\_id'. The figure demonstrates the difficulties that arise when identifying matching data, even within a database that shares an identical schema.

\subsection{Experimental Setup}
In the following section, we detail the experimental setup used to evaluate our approach. This includes a comprehensive description of the training and testing dataset (Section \ref{sec:dataset}), the evaluation metrics employed (Section \ref{sec:evaluation-metrics}), the validation process for the LLM model utilized in the training data generation (Section \ref{sec:verify-llm}), the details on training the PLM (Section \ref{sec:baseline}), and the details on the baseline methods (Section \ref{sec:baseline-info}).

\subsubsection{Dataset}\label{sec:dataset}
We use the Tungsten assessment dataset~\cite{tungsten-assessment}, a manually curated dataset by the USGS, to train one of the baseline models (detailed in Section \ref{sec:baseline-info}) and to evaluate the performance of both the baseline models and proposed approach. This dataset contains records from the MRDS and USMIN databases focusing on Tungsten mining sites in Idaho and Montana. The Tungsten assessment dataset comprises 387 records (383 from MRDS and 4 from USMIN), generating a total of 74,691 potential record pairs for analysis. The count of \textit{match} pairs and the count of \textit{non-match} pairs are shown in Table \ref{tab:dataset_detail}.

To train our approach, we generate synthetic training data by asking an LLM to label ``Yes'' (later mapped to \textit{match}) and ``No'' (later mapped to \textit{non-match}) on randomly selected MRDS and USMIN records. We ensure that the volume of data--387 records and 74,691 record pairs--is identical to that of the Tungsten assessment data. We randomly select records from the entire dataset, which leads to the data covering diverse minerals (e.g., Zinc and Copper) and a broader geographical range (i.e., across the United States). 

Additionally, a USGS expert has curated a Nickel ground truth data\footnote{Michael Zientek from USGS has verified the accuracy of the data.} covering the Upper Midwest United States. We use this dataset exclusively to evaluate the baseline models and our proposed approach. The dataset comprises 24 mineral site records, resulting in 276 potential record pairs for model evaluation.

Details of the datasets are available in Table \ref{tab:dataset_detail}.

\begin{table}[htp!]
    \centering
    \begin{tabular}{|c||c|c|}
    \hline
         & \# Match Pairs & \# Non-Match Pairs \\
        \hline \hline
       Ground Truth (Tungsten) & 23 & 74668 \\
       Ground Truth (Nickel) & 18 & 258 \\
       LLM Labeled & 437 & 74254 \\
       \hline
    \end{tabular}
    \caption{Count of match pairs and non-match pairs in each dataset.}
    \label{tab:dataset_detail}
\end{table}

\subsubsection{Evaluation Metrics} \label{sec:evaluation-metrics}




To evaluate the performance of our approach in classifying record pairs into each category, we measure three types of F1 scores: the \textit{match} class F1 score (Equation \ref{eq:match}), the \textit{non-match} class F1 score (Equation \ref{eq:nomatch}), and the macro-averaged F1 score (Equation \ref{eq:macro}). In these equations, $tp$ represents the count of true positives, $tn$ represents true negatives, $fp$ represents false positives, and $fn$ represents false negatives.

\begin{equation} \label{eq:match}
    \text{Match F1} = \frac{2tp}{2tp+fp+fn}
\end{equation}

\begin{equation} \label{eq:nomatch}
    \text{Non-match F1} = \frac{2tn}{2tn+fp+fn}
\end{equation}

\begin{equation} \label{eq:macro}
    \text{Macro-averaged F1} = \frac{tp}{2tp+fp+fn} + \frac{tn}{2tn+fp+fn}
\end{equation}

We choose the macro-averaged F1 score over other average F1 score variants to account for the class imbalance~\cite{f1score} in the test dataset.

\subsubsection{LLM Performance Validation} \label{sec:verify-llm}
Before generating the labeled training data, we first validate the performance of the LLaMA3-8b~\cite{dubey2024llama3herdmodels} model on the complete Tungsten and Nickel datasets, without partitioning the data into training, validation, and testing sets. For the 74,691 Tungsten record pairs, LLaMA has a match F1 score of 39.13\% and a non-match F1 score of 99.96\%. For the 276 Nickel pairs, LLaMA has a match F1 score of 70.96\% and a non-match F1 score of 98.27\%. Table \ref{tab:llama-performance} provides a breakdown of the number of correctly identified \textit{match} and \textit{non-match} pairs in relation to the total number of pairs available in the ground truth data.

\begin{table*}[hbt!]
    \centering
    \begin{tabular}{|c||c|c|c|c|}
        \hline
        & \multicolumn{2}{c|}{Matches} & \multicolumn{2}{c|}{Non-matches} \\
        \hline
        & Correctly Identified & Total Pairs & Correctly Identified & Total Pairs \\
        \hline \hline
        Ground Truth (Tungsten) & 18 & 23 & 74,617 & 74,668 \\
        Ground Truth (Nickel) & 11 & 18 & 256 & 258 \\
        \hline
    \end{tabular}
    \caption{Performance of the LLaMA3-8b model on the record linkage task, showing the number of correctly identified match pairs compared to the total match pairs in the ground truth, as well as the number of correctly identified non-match pairs compared to the total non-match pairs in the ground truth.}
    \label{tab:llama-performance}
\end{table*}

However, the inference time was substantial at approximately 3 hours on our computation resource (detailed in Section \ref{sec:computation-resource}). The computation involves $_{n}C_{2}$ comparisons, where $n$ represents the number of records. As a result, the number of combinations grows quadratically with the number of records, leading to a corresponding quadratic increase in time complexity. This highlights the impracticality of relying solely on LLMs for large-scale record linkage tasks, particularly in scenarios involving extensive datasets. Mineral site databases such as MRDS and USMIN are archives where sites have been recorded over decades and are represented by multiple records across various databases. For instance, the MRDS~\cite{mrds} contains over 300,000 records, and the USMIN database continues to expand~\cite{usmin}. The challenges regarding efficiency and scalability need to be addressed. To verify that our approach is more efficient compared to LLM-relying record linkage models, we compare the inference time based purely on LLM and our approach in relation to the number of records.


\subsubsection{Fine-Tuning PLMs}\label{sec:baseline}
We randomly select 80\% of the LLaMA-labeled data to train a RoBERTa model~\cite{roberta} and evaluate its performance using 10\% of the Tungsten ground truth data as the test set. We partition the data such that each subset contains a proportionate number of \textit{match} and \textit{non-match} samples, maintaining a balance across all data partitions. The model is trained for 10 epochs, and the best-performing epoch, as determined by the validation data, is selected for final testing. The model uses a batch size of 32, a learning rate of 2e-5, and a weight decay of 0.01\footnote{All other parameters not explicitly mentioned follow the default configuration provided by the RoBERTa model available on \href{http://huggingface.co/docs/transformers/en/model_doc/roberta}{HuggingFace RoBERTa page}.}.

\subsubsection{Baseline Methods}\label{sec:baseline-info}
To evaluate the effectiveness of our approach, we compare it against four baseline methods: GT-Trained, Manually Curated, LLaMA3-8b, and GeoER:

\begin{itemize}
    \item \textbf{GT-Trained}: This method uses 80\% of the Tungsten ground truth (GT) data for training, 10\% for validation, and 10\% for testing. Similar to our approach, we partition the data so that each data partition contains a proportionate number of \textit{match} and \textit{non-match} samples. We fine-tune a RoBERTa model using the same hyperparameter settings as our approach to ensure a fair comparison.
    
    \item \textbf{Manually Curated}: This method utilizes empirically defined distance thresholds and text similarity scores. We select parameters that yield the highest performance on the training and validation datasets, which consist of 90\% of the Tungsten ground truth data. This method achieves optimal results with a 5-kilometer distance threshold and an 85\% cosine similarity score for text embeddings generated using Sentence-BERT~\cite{sbert}.
    
    \item \textbf{LLaMA3-8b}: This method relies exclusively on LLMs, specifically the LLaMA3-8b model, for mineral site record linkage. It does not incorporate spatial information or predefined thresholds.
    
    \item \textbf{GeoER}~\cite{10.1145/3485447.3512026}: This method serves as a baseline method for spatial record linkage. To implement the GeoER framework, we reformat our data according to the authors' guidelines to ensure compatibility with their experimental setup. We maintain the original parameters defined by the authors\footnote{Details on the model are available on the \href{https://github.com/PasqualeTurin/Geo-ER}{GeoER page}}.
\end{itemize}





\subsubsection{Hardware Details} \label{sec:computation-resource}
We perform all experiments on one 40GB A100 GPU chip and dedicate 70GB of memory at most; the majority of the memory is used to load and use the LLaMA3-8b model\footnote{Specific details about the model, such as model architecture and training data, are available on \href{https://huggingface.co/meta-llama/Meta-Llama-3-8B}{Huggingface meta-llama/Meta-Llama-3-8B page}.} for training data generation. 

\subsection{Experimental Results} \label{sec:experimental-results}
Table \ref{tab:results} displays the match, non-match, and macro-averaged F1 scores of the baseline models and our proposed approach. All values are reported in percentage (\%).

\begin{table*}[hbt!]
    \centering
    \begin{tabular}{|c||c|c|c|c|c|c|c|}
    \hline
         & \multicolumn{3}{c|}{Tungsten Data} & \multicolumn{3}{c|}{Nickel Data} \\
         \hline
         & Match F1 & Non-match F1 & Macro-avg F1 & Match F1 & Non-match F1 & Macro-avg F1\\
         \hline \hline
        GT-Trained & 0.00 & \textbf{99.98} & 49.99 & 0.00 & 96.63 & 48.31 \\
        Manually Curated & 28.57 & \underline{99.97} & 64.27 & 50.00 & \underline{97.73} & 73.86 \\
        LLaMA3-8b & \textbf{46.15} & 99.95 & \textbf{73.05} & \textbf{77.78} & \textbf{98.45} & \textbf{88.11}\\
        GeoER~\cite{10.1145/3485447.3512026} & - & - & - & 22.22 & 95.10 & 58.66 \\
        \hline
        Our Approach & \textbf{46.15} & 99.95 & \textbf{73.05} & \underline{57.14} & 97.71 & \underline{77.43}\\
        \hline
    \end{tabular}
    \caption{F1 score of match class, F1 score of non-match class, and macro-averaged F1 score of each model on the Tungsten and Nickel dataset. The highest scores are bold, and the second highest are underlined. Models with missing data were not completed within the 23-hour time limit.}
    \label{tab:results}
\end{table*}

As shown in Table \ref{tab:results}, our approach outperforms the GT-Trained model in both the match F1 score and macro-averaged F1 score for both the Tungsten and Nickel datasets. The GT-Trained model predicts ``No'' for all record pairs due to the lack of \textit{match} samples in the ground truth training data, while our approach provides more balanced predictions across both categories. Additionally, our approach performs at the same level as the LLaMA3-8b model for the Tungsten dataset but lower for the Nickel dataset. We explain the benefit of our model compared to the LLaMA3-8b model in Section \ref{sec:run-time}.

Despite GeoER~\cite{10.1145/3485447.3512026} using spatial aspects in its training and inference process, it still underperforms compared to our model, which does not explicitly incorporate spatial considerations. Additionally, GeoER cannot complete the training on the Tungsten dataset within the 23-hour timeframe. To provide context, the POI dataset originally used by GeoER consists of 2,500 records, and GeoER takes more than 10 hours to train. In contrast, our Tungsten training dataset consists of 59,752 records, more than 23 times the size of the POI dataset. Given the increase in data volume, GeoER's training time becomes excessively long, making it impractical for large-scale mineral site linkage tasks. With the future incorporation of spatial consideration into our model, we anticipate higher performance beyond what is currently shown in Table \ref{tab:results}. This further solidifies our claim that our approach is a more robust and scalable solution for mineral site record linkage. 

\begin{figure*}
    \centering
    \includegraphics[width=0.9\linewidth]{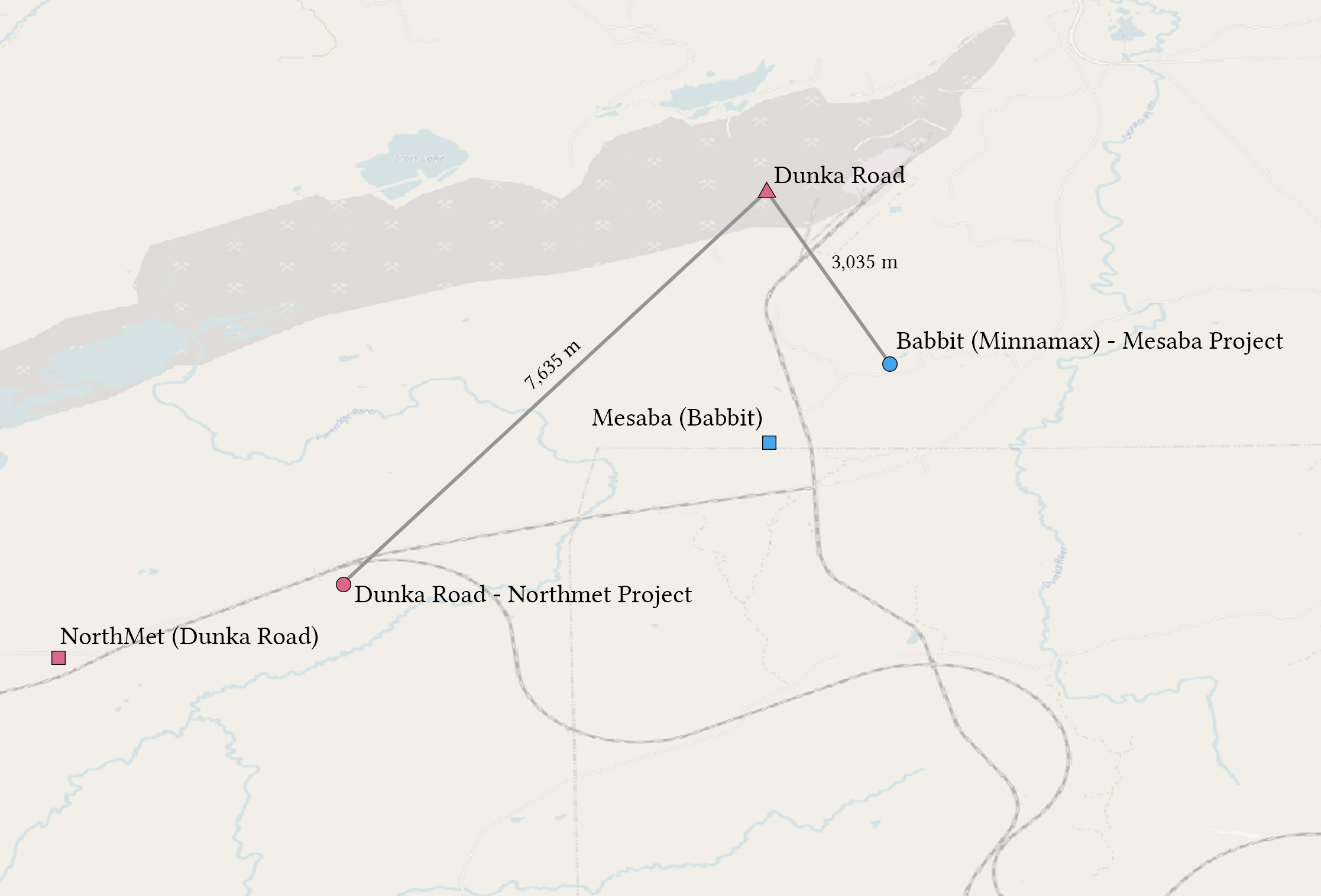}
    \caption{Image illustrating true positive match records identified by our proposed approach. Each color represents a ground-truth mineral site, and each shape corresponds to a different database. Although `Dunka Road' is geographically closer to `Babbit (Minnamax - Mesaba Project)', our approach distinguishes them as separate mines. In contrast, it identifies `Dunka Road' and `Northmet Project' located at a farther distance as belonging to the same mineral site.}
    \label{fig:result-example}
\end{figure*}

The qualitative results of our proposed approach demonstrate the characteristics of an effective record linkage model. Trained on LLaMA-labeled data, our model accurately classifies pairs or records as \textit{match} and \textit{non-match}. As illustrated in Figure \ref{fig:result-example}, our model successfully links `Dunka Road' with `Dunka Road - Northmet Project', even though the sites are over 7 kilometers apart. Additionally, the model correctly identifies `Dunka Road' and `Babbit (Minnamax) - Mesaba Project' as distinct mineral sites despite their close proximity (3 kilometers) by recognizing other differentiating characteristics.

\subsubsection{Types of Error} \label{sec:types-of-error}

In both the Tungsten and Nickel datasets, our approach encounters challenges due to unclear site distinctions. For example, our approach struggles to differentiate two sites named `Unidentified Occurrence', treating them as identical sites. However, determining whether such cases should be classified as \textit{match} or \textit{non-match} is a complex design challenge (not a model error) in both manual and automated scenarios. 

Unlike the Tungsten dataset, the Nickel dataset contains uncleaned data, making the record linkage process more challenging. The Nickel dataset includes entries for large deposit regions (e.g., `Duluth Complex') and records for multiple individual sites (e.g., `Maturi, Birch Lake, and Spruce Road Copper-Nickel'). To illustrate the spatial coverage of these large deposits, Figure \ref{fig:duluth-complex} shows the geographical spread of the `Duluth Complex' in Minnesota. Since the `Duluth Complex' encompasses multiple mineral sites, a practical model should exclude data representing a general region (e.g., `Duluth Complex') when performing record linkage. 

The current pipeline requires a one-to-one comparison between records, which leads to not capturing some links. In some cases, information needs to be inferred from other records, yet the current structure prevents such. For example, a record in the database may list `Dunka Pit' as a potential alternative name for the `Northmet Project.' However, in a different dataset, this alternative name is not recorded. When the model compares this record with another that lists `Dunka Pit' as the primary name, it fails to identify them as the same site, as it does not account for information found in out-of-pair records. We can potentially address this limitation by constructing a graph-like structure from the data. By treating each record as a node and each labeled match as a connecting edge, we can establish links across multiple records. This approach would allow us to link sites, such as the three aforementioned mines, based on their indirect connections through shared information.

\subsubsection{Inferences Time} \label{sec:run-time}
\begin{figure}[htp!]
    \centering
    \includegraphics[width=1\linewidth]{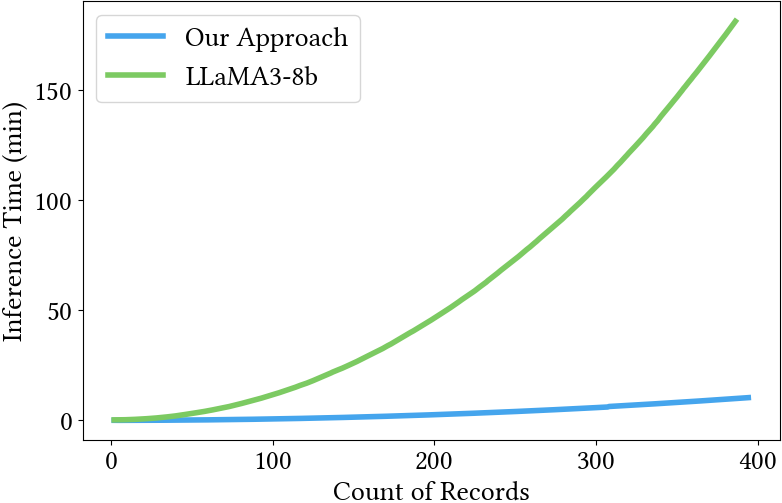}
    \caption{Inference time (in minutes) for our approach compared to the LLaMA-relying model, shown in relation to the number of mineral site records. Our approach is represented in blue, while the LLaMA model is depicted in green. The LLaMA-relying model demonstrates significantly longer inference times compared to our approach.}
    \label{fig:run-time}
\end{figure}

We compare the inference time of our approach against the LLaMA3-relying model. We measure the inference time by incrementing the testing size from 10 mineral site records (i.e., 45 pairs) to 300 mineral site records (i.e., 44,850 pairs). Figure \ref{fig:run-time} displays the inference time of each model in relation to the number of mineral site records. The green line shows the inference time when purely relying on LLaMA3-8b, and the blue line shows the inference time when using our approach. The time required for inference is lower for ours than relying on LLaMA. With the collected inference time, we derive an equation to estimate the required inference time for running it at the scope of the actual mineral site linkage task. Details are provided in Appendix \ref{sec:estimate-runtime}.

\subsubsection{Training Data Size} \label{sec:data-size}
\citet{sample-size} states that BERT models may not predict a specific label until the number of training samples exceeds a threshold. To better understand the relationship between data size, level of class imbalance, and performance, we perform additional experiments by varying the training dataset size. In all experiments, we use LLM-labeled data to fine-tune the RoBERTa model and evaluate the model using the Tungsten ground truth dataset.

\begin{figure}[htp!]
    \centering
    \includegraphics[width=1\linewidth]{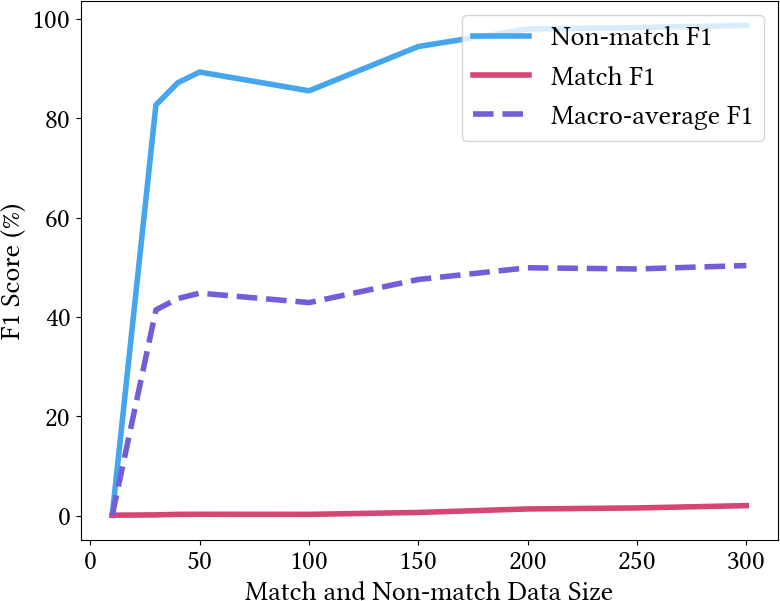}
    \caption{Match, non-match, and macro-averaged F1 scores of our model, plotted against the size of the match and non-match training samples. The `match and non-match data size' represents the number of training samples per class. While non-match F1 and macro-averaged F1 scores stabilize at 150 samples per class, match F1 shows only marginal improvement throughout the experiment.}
    \label{fig:data-size}
\end{figure}

We first increase the number of \textit{match} and \textit{non-match} samples at the same rate, increasing from 10 samples in each class to 300 samples in each class. Figure \ref{fig:data-size} shows the match, non-match, and macro-averaged F1 score for each training sample size. Though the F1 score for each class increases, the increment for match F1 is minimal. 

\begin{figure}[htp!]
    \centering
    \includegraphics[width=1\linewidth]{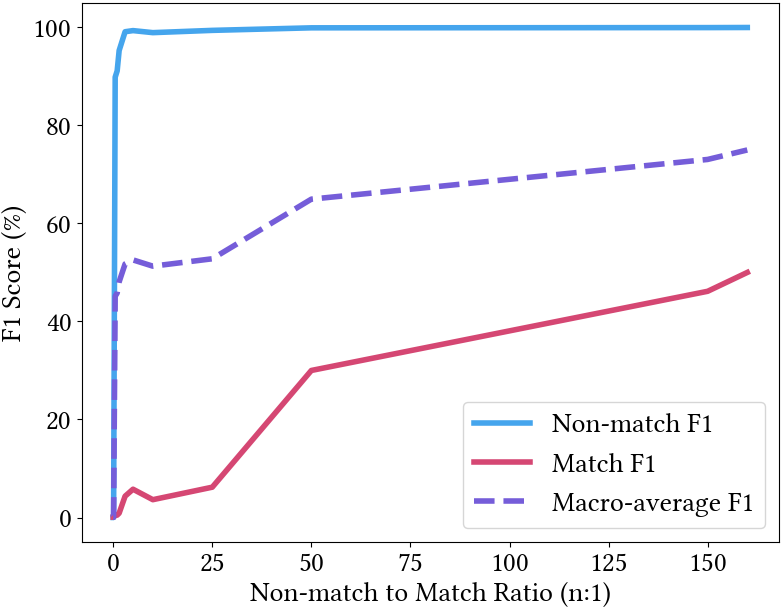}
    \caption{Match, non-match, and macro-averaged F1 scores of our model, plotted against the ratio of non-match to match training samples. The number of match samples is fixed at 349, while the non-match sample size increases according to the specified ratio. The non-match F1 score stabilizes when the non-match sample size is approximately 10 times that of the match samples, whereas both the match F1 and macro-averaged F1 scores continue to improve as the ratio increases.}
    \label{fig:data-ratio}
\end{figure}

We further the experiment by fixing the number of \textit{match} samples at 349 and increasing the number of \textit{non-match} samples from 0 to 160 times the match sample size. Figure \ref{fig:data-ratio} illustrates that the non-match F1 experiences a significant increase until it reaches a saturation point, while the match F1 score continuously rises with the increasing ratio of \textit{non-match} data. Despite the growing imbalance in the training data, the overall performance continues to improve as the total data size increases. This implies that beyond a certain single-class data size threshold, the total data size determines the model's overall performance.

\begin{figure}[htp!]
    \centering
    \includegraphics[width=1\linewidth]{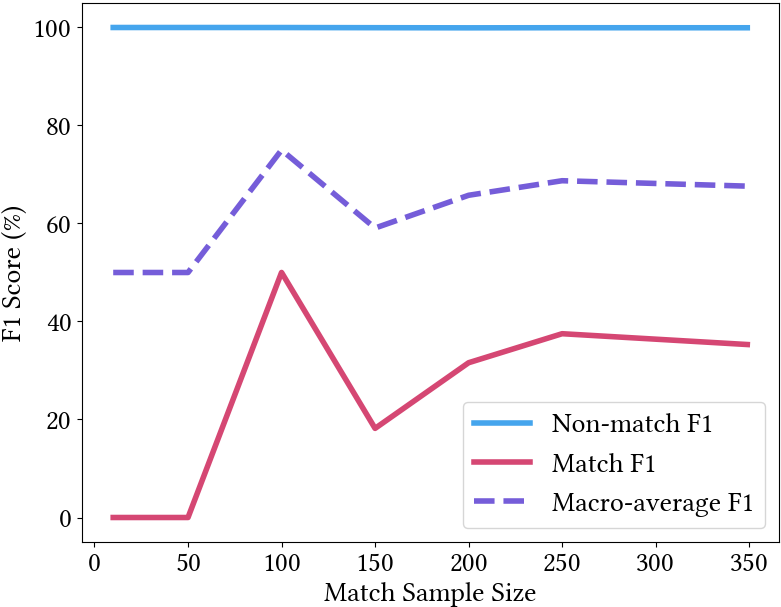}
    \caption{Match, non-match, and macro-averaged F1 scores of our model, plotted against the size of the match training samples. The number of non-match samples is fixed at 59,403, while the match sample size increases from 10 to 350. The non-match F1 score remains nearly constant, while the match F1 and macro-averaged F1 scores peak at a match sample size of 100, decrease, and then increase again, stabilizing around 250 samples.}
    \label{fig:yes-size}
\end{figure}

To determine the training data size threshold for a single class, we conduct an additional experiment by varying the number of \textit{match} samples while keeping the number of \textit{non-match} samples constant at 59,403 pairs. As presented in Figure \ref{fig:yes-size}, the experiment reveals an interesting pattern with the F1 score peaking at 100 \textit{match} samples and stabilizing after 250 \textit{match} samples. This finding supports our claim that beyond a certain threshold of minority class samples (in this case, the \textit{match} samples), the overall data size drives the model performance. This indicates that the model learns key patterns in the \textit{match} samples through exposure to \textit{non-match} samples.

\section{Discussion and Future Work}
Linking mineral site records is an essential task for mineral deposit prospecting and resource management~\cite{prospectivity}. These records are often sourced from structured databases like MRDS and USMIN, but they can also be extracted from reports, webpages, or other unstructured data. The complexity of this task stems from several factors, including ambiguous or incomplete location data, inconsistent site naming conventions, and unreliable geographical distances. In this paper, we propose a method leveraging LLMs to generate training data for mineral site record linkage, addressing the challenges of limited ground-truth data. Our approach demonstrates the potential of using LLM-generated data to fine-tune PLMs, achieving improved performance compared to models trained on imbalanced ground-truth datasets. Future work will focus on refining our proposed method and addressing two key questions to enhance model performance further:

\subsection{How can spatial data be more effectively integrated into the pipeline?}
At the current stage, spatial data has been treated as part of the input string to the PLM without special handling; the main goal of the research is to demonstrate the potential of leveraging LLMs to generate training data to fine-tune PLMs. However, we recognize the importance of preserving the spatial semantics in the data. Moving forward, we plan to explore more effective ways to incorporate spatial attributes into the PLM to enhance its accuracy in handling the spatial dimension of such data. 

We propose three potential approaches for this. We can utilize spatial distance between records, as demonstrated by \citet{10.1145/3485447.3512026}, using haversine distance to create distance embeddings concatenated with the serialized record pair embedding. Alternatively, the spatial distance can be embedded in each serialized entity using a spatial coordinate embedding, as suggested in SpaBERT~\cite{spabert}. For the third approach, we can incorporate spatial location as an additional attribute for each record. We can convert the latitude and longitude values into point embeddings and concatenate them to each serialized entity embeddings.

\subsection{Can LLMs generate sufficient data for training?}
We have observed performance improvements with the increased number of \textit{match} samples (i.e., positive samples) in the training data. The current data generation process randomly selects potential record pairs without ensuring a sufficient number of match samples. Therefore, to gather a sufficient number of training data, we need to execute the LLaMA labeling pipeline iteratively until the training data includes an adequate number of match samples. 

A logical next step is to investigate whether increasing the number of \textit{match} samples leads to additional performance gains. Previous research~\cite{borisov2023language} demonstrates that LLMs can generate synthetic tabular data. We aim to explore whether LLMs can synthesize entirely new data that replicate the patterns of \textit{match} and \textit{non-match} records in the existing dataset. This could balance the representation of match samples and further improve the performance of pre-trained language models (PLMs) when trained on LLM-generated data.

By combining the efficiency of PLMs with the advanced data generation capabilities of LLMs, we have demonstrated that our approach offers a scalable and automated solution to these challenges. Our results show that LLM-generated data can significantly reduce the need for manual data curation and support more accurate and efficient record linkage across large-scale datasets. Moving forward, we will continue to enhance our method to fully understand the potential of LLMs in addressing complex real-world data challenges with greater efficiency and accuracy.




\section*{Ethical Consideration}
We acknowledge that the proposed approach allows third-party LLMs to access the data to generate training datasets. 

While our approach enhances the efficiency of the record linkage process, it is essential to recognize that using LLMs does incur energy consumption and contributes to carbon emissions. However, our method minimizes these environmental costs by utilizing LLMs only during the training data generation phase rather than relying on LLMs throughout the entire record linkage process. 

Additionally, LLMs may generate data that deviates from the ground truth pattern. These hallucinations pose risks to PLMs trained on such data, potentially causing erroneous outcomes. Careful validation of LLM-generated data is essential to prevent the propagation of inaccurate information. To apply the proposed approach in practical record linkage applications, robust safeguards, and error detection mechanisms will be essential to ensure reliability and accuracy.

\section*{Acknowledgments}
This material is based upon works supported by the Defense Advanced Research Projects Agency (DARPA) under Agreement No. HR00112390132 and Contract No. 140D0423C0093. Any opinions, findings and conclusions or recommendations expressed in this material are those of the authors and do not necessarily reflect the views of the Defense Advanced Research Projects Agency (DARPA); or its Contracting Agent, the U.S. Department of the Interior, Interior Business Center, Acquisition Services Directorate, Division V.

We extend our gratitude to Dr. Graham W. Lederer (USGS) for his valuable geological insights, Dr. Michael L. Zientek (USGS) for his assistance in generating the Nickel ground truth data, and Margaret A. Goldman (USGS) for sharing her expertise and experience in creating manually linked ground truth data.

\bibliographystyle{ACM-Reference-Format}
\bibliography{main}

\onecolumn \newpage
\appendix
\section{Span of Duluth Complex}
\begin{figure*}[htp!]
    \centering
    \includegraphics[width=0.8\linewidth]{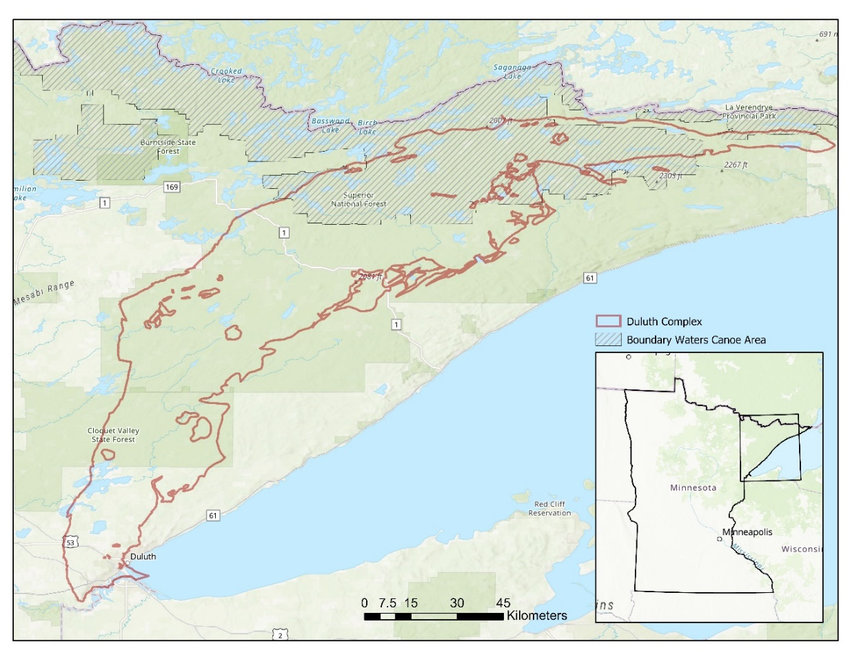}
    \caption{Map of the Duluth Complex. Image sourced from~\citet{duluth-complex}.}
    \label{fig:duluth-complex}
\end{figure*}

Figure \ref{fig:duluth-complex} illustrates the extensive spatial coverage of the Duluth Complex in Minnesota, outlined by a red boundary. The area of the complex exceeds 5,000 $km^{2}$~\cite{duluth-complex-information}. An effective mineral site record linkage model should avoid linking large region records to smaller mineral site records.

\section{Performance Comparison between BERT Variants} \label{sec:different-bert}
We select the RoBERTa model over BERT and its variants due to its well-known performance across various downstream tasks~\cite{roberta}, along with the relatively low training and inference times. We conduct additional experiments to show that the RoBERTa model outperforms other variants when using our approach for the classification task.

\begin{table*}[hbt!]
    \centering
    \begin{tabular}{|c||c|c|c|}
        \hline
         & Match F1 & Non-match F1 & Macro-avg F1 \\
         \hline\hline
        BERT-cased & 0.00 & \underline{99.98} & 49.99 \\
        BERT-uncased & \underline{17.65} & 99.81 & \underline{58.73} \\
        DeBERTa & 7.14 & 99.65 & 53.40 \\
        \hline
        RoBERTa (Selected Model) & \textbf{46.15} & \textbf{99.95} & \textbf{73.05} \\
        \hline
    \end{tabular}
    \caption{Match F1, Non-match F1, and Macro-averaged F1 scores of BERT-cased, BERT-uncased, DeBERTa, and RoBERTa fine-tuned using our approach. The highest scores are highlighted in bold, with the second-highest scores underlined.}
    \label{tab:bert-variants}
\end{table*}

We compare the performance of RoBERTa against BERT-cased, BERT-uncased, and DeBERTa, with all models fine-tuned using the same LLaMA-generated data and tested on the same Tungsten data. We configure all four models with a batch size of 32, a learning rate of 2e-5, and a weight decay of 0.01\footnote{All other parameters we do not explicitly mention follow each model's default configuration. These parameters are available here: \href{https://huggingface.co/google-bert/bert-base-cased}{HuggingFace BERT-cased}, \href{https://huggingface.co/google-bert/bert-base-uncased}{HuggingFace BERT-uncased},  \href{https://huggingface.co/microsoft/deberta-v3-base}{HuggingFace DeBERTa}, \href{https://huggingface.co/FacebookAI/roberta-base}{Huggingface RoBERTa}.}. Table \ref{tab:bert-variants} shows the classification performance of each model. The result indicates that fine-tuning RoBERTa is best suited for our approach, and, therefore, we conduct further analyses (i.e., data-size comparison and run-time comparison) using the fine-tuned RoBERTa model.

\section{Runtime Estimation for Full Mineral Site Data} \label{sec:estimate-runtime}
Using the recorded inference times, we extrapolate the data to estimate the time required to compare 300,000 records, which reflects the scope of the mineral site linkage task. To extrapolate, we derive a quadratic function that relates inference time ($time$) to the number of records ($count_{ms}$). 

For the LLaMA model, we approximate the inference time (in seconds) with the following equation:
\begin{equation} \label{eq:llama-graph}
time = 0.073 \cdot (count_{ms}^2 - count_{ms})
\end{equation}

For our approach, we approximate the inference time (in seconds) with the following equation:
\begin{equation} \label{eq:our-graph}
time = 0.004 \cdot (count_{ms}^2 - count_{ms})
\end{equation}

Based on these calculations, our approach can complete the task in approximately 4,166 days, whereas relying solely on the LLaMA model will require around 76,000 days. However, we can filter out many candidate pairs through basic spatial comparisons. For example, a mineral site in the central United States is unlikely to be linked with a mineral site in the southeastern United States; we can eliminate these pairs before running the model. As a result, the actual inference time will be lower than the estimated values.

\end{document}